\documentclass[reqno]{amsart}

\usepackage[dvips]{graphicx}

\begin{document}

\title[Solitary wave solutions of the SPE]{Solitary wave solutions\\ of the short pulse equation}

\author{Anton Sakovich}
\address{Anton Sakovich \newline Department of Physics, Belarusian State University, 220080 Minsk, Belarus}
\email{ant.s@tut.by}

\author{Sergei Sakovich}
\address{Sergei Sakovich \newline Institute of Physics, National Academy of Sciences, 220072 Minsk, Belarus}
\email{saks@tut.by}

\begin{abstract}
An exact nonsingular solitary wave solution of the Sch\"{a}fer--Wayne short pulse equation is derived from the breather solution of the sine-Gordon equation by means of a transformation between these two integrable equations.
\end{abstract}

\maketitle

The short pulse equation (SPE), which has the form
\begin{equation}
	u_{xt} = u + \tfrac{1}{6} \left( u^3 \right)_{xx}
	\label{spe}
\end{equation}
up to scale transformations of its variables, was derived by Sch\"{a}fer and Wayne \cite{SW} as a model equation describing the propagation of ultra-short light pulses in silica optical fibers. In contrast to the nonlinear Schr\"{o}dinger equation (NLSE) which models slowly varying wave trains, the SPE approximates Maxwell's equations in the case when the pulse spectrum is not narrowly localized around the carrier frequency, that is when the pulse is as short as a few cycles of the central frequency. Numerical simulations made in \cite{CJSW} showed that the accuracy of the SPE approximation to Maxwell's equations steadily increases as the pulse shortens, whereas the NLSE approximation fails to be accurate for ultra-short pulses.

Like the NLSE, the SPE is an integrable nonlinear partial differential equation. Actually, the equation \eqref{spe} appeared first a long time ago, as one of Rabelo's equations possessing a zero-curvature representation with a parameter \cite{R} (we are grateful to Prof.\ E.G. Reyes for this essential reference; see also \cite{BRT}). Recently, the Lax pair of the SPE, of the Wadati--Konno--Ichikawa (WKI) type, was rediscovered in \cite{SS}, where also the second-order recursion operator of the SPE was presented and the chain of transformations
\begin{equation}
	\begin{split}
	v & (x,t) = \left( u_x^2 + 1 \right)^{-1/2} ; \\
	x = w & (y,t) , \quad v(x,t) = w_y (y,t) ; \\
	z & (y,t) = \arccos w_y
	\end{split}
	\label{chain}
\end{equation}
was constructed which relates the SPE \eqref{spe} with the sine-Gordon equation (SGE)
\begin{equation}
	z_{yt} = \sin z .
	\label{sge}
\end{equation}
Very recently, the bi-Hamiltonian structure of the SPE was discovered and the corresponding hierarchy and conservation laws were studied in \cite{B1,B2}.

The soliton solutions of the NLSE have played an important role in the recent development of fiber-optic communications. For this reason, one would definitely like to have exact mathematical expressions for the ultra-short light pulses governed by the SPE. In the present paper, we derive exact solitary wave solutions of the SPE \eqref{spe} from the well-known ones of the SGE \eqref{sge} by means of the transformation \eqref{chain} between these two integrable equations.

Sch\"{a}fer and Wayne \cite{SW} proved the nonexistence theorem that the SPE does not possess any solution representing a smooth localized pulse moving with constant shape and speed. Actually, it is easy to find the ``wave of translation'' solution of the SPE directly, in order to see that this solution is singular. Substituting
\begin{equation}
	u = u(\xi) , \quad \xi = x + c t , \quad c = \mathrm{constant}
	\label{wave}
\end{equation}
into \eqref{spe}, we get the second-order ordinary differential equation
\begin{equation}
	\left( u^2 - 2 c \right) u_{\xi \xi} + 2 u \left( u_{\xi}^2 + 1 \right) =0 .
	\label{2ord}
\end{equation}
Then, using the integrating factor $2 \left( u^2 - 2 c \right) u_{\xi}$, we get from \eqref{2ord} the first-order equation
\begin{equation}
	\left( u^2 - 2 c \right)^2 \left( u_{\xi}^2 + 1 \right) = d ,
	\label{1ord}
\end{equation}
where we must set $d = 4 c^2$ for the constant of integration $d$, in order that the asymptotic conditions $| u | \to 0$ and $| u_{\xi} | \to 0$ at $| \xi | \to \infty$ be satisfied. The equation \eqref{1ord} in its form
\begin{equation}
	u_{\xi} = \pm \dfrac{u \sqrt{4 c - u^2}}{2 c - u^2}
	\label{sing}
\end{equation}
clearly shows that the sought solution $u$ inevitably reaches the value $u = \sqrt{2 c}$ or $u = - \sqrt{2 c}$, where the solution has the singularity $u_{\xi} \to \pm \infty$. Further integration of the equation \eqref{sing} is possible, and it would bring us to the loop soliton solution of the SPE \eqref{spe}. We will obtain this solution soon, but in a different way.

For simplicity of the subsequent practical use, we can rewrite the transformation \eqref{chain} in the following equivalent form:
\begin{gather}
	u(x,t) = z_t (y,t) ,
	\label{tru} \\
	x = w(y,t) : \quad \left\{
	\begin{aligned}
		w_y & = \cos z , \\
		w_t & = - \tfrac{1}{2} z_t^2 .
	\end{aligned}
	\right.
	\label{trx}
\end{gather}
The system of two equations for $w(y,t)$ inside \eqref{trx} is compatible if and only if $z(y,t)$ satisfies the SGE \eqref{sge}. For any solution $z$ of the SGE \eqref{sge}, the relation \eqref{tru} uniquely determines $u$ as a function of $y$ and $t$, and the relation \eqref{trx} determines $x$ as a function of $y$ and $t$ up to an additive constant of integration. This yields a solution $u(x,t)$ of the SPE \eqref{spe}, given in a parametric form, with $y$ being the parameter. Thus, using the transformation \eqref{tru}--\eqref{trx}, we can easily derive exact solutions of the SPE from the known ones of the SGE. Note that the invariance of the SGE \eqref{sge} under the Lorentz transformation
\begin{equation}
	y \mapsto a y , \quad t \mapsto a^{-1} t , \quad z \mapsto z
	\label{ltr}
\end{equation}
corresponds via \eqref{tru}--\eqref{trx} to the invariance of the SPE \eqref{spe} under the scale transformation
\begin{equation}
	x \mapsto a x , \quad t \mapsto a^{-1} t , \quad u \mapsto a u ,
	\label{str}
\end{equation}
where $a = \mathrm{constant} \neq 0$. Therefore we can put the source solution of the SGE into a simpler form by \eqref{ltr}, thus simplifying the symbolic integration required in \eqref{trx}, and then use \eqref{str} to generalize the target solution of the SPE. Also we can simplify the source solution of the SGE by shifts of $y$ and $t$, $y \mapsto y + y_0$ and $t \mapsto t + t_0$: a shift of $t$ in $z(y,t)$ causes the same shift of $t$ in $u(x,t)$, while a shift of $y$ has no effect on the target solution of the SPE.

\begin{figure}
	\includegraphics[width=6cm]{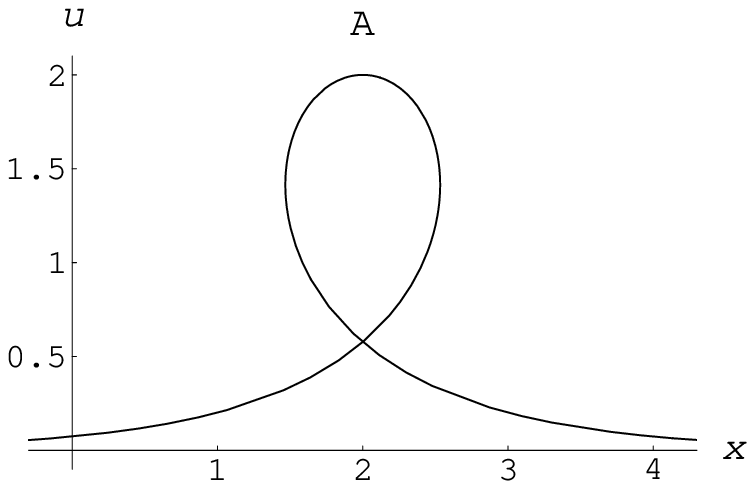}
	\includegraphics[width=6cm]{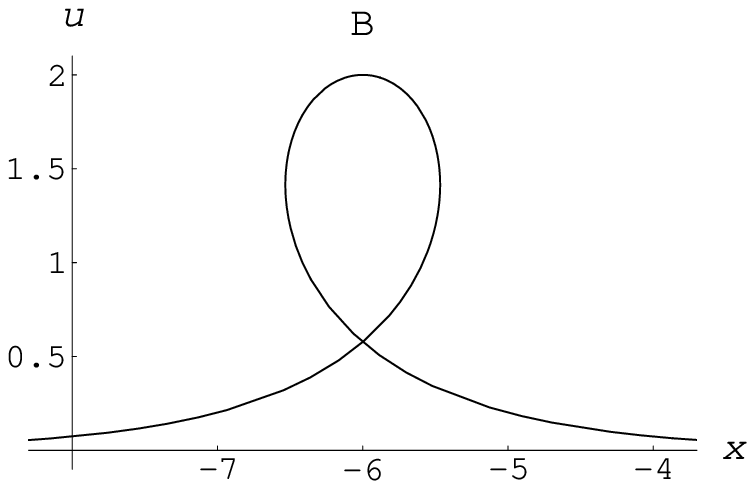}
	\caption{The loop soliton solution $u(x,t)$ \eqref{loop} of the SPE \eqref{spe}: (A) $t = -2$; (B) $t = 6$.}
	\label{fig1}
\end{figure}

Now let us see how the transformation \eqref{tru}--\eqref{trx} works. It is natural to begin with the kink solution of the SGE \eqref{sge} \cite{L}, in the form
\begin{equation}
	z = 4 \arctan \bigl( \exp ( y + t ) \bigr)
	\label{kink}
\end{equation}
simplified by a Lorentz transformation and a shift of $y$. In the case of $z(y,t)$ given by \eqref{kink}, the relations \eqref{tru} and \eqref{trx} immediately yield
\begin{equation}
	u = 2 \operatorname{sech} ( y + t ) , \quad x = y - 2 \tanh ( y + t ) ,
	\label{loop}
\end{equation}
where we have fixed the constant of integration in $x$ so that $x|_{y=t=0} = 0$. The obtained solution $u(x,t)$ \eqref{loop} of the SPE \eqref{spe} is the loop soliton which moves from the right to the left with constant shape and unit speed, as shown in Figure~\ref{fig1}.
This multi-valued solution satisfies the equation \eqref{sing} with $c = 1$ and has two singularities with $u_x = u_t \to \pm \infty$ at the level $u = \sqrt{2}$. (Note that, if we chose the antikink solution of the SGE as a source solution for the transformation \eqref{tru}--\eqref{trx}, the target solution of the SPE would be the antiloop soliton which differs from the loop soliton in the sign of $u$ only.) We can generalize the solution \eqref{loop} by the transformation \eqref{str}, thus obtaining either a bigger and faster loop or a smaller and slower one. Of course, we could expect that the SPE possesses a loop soliton solution, since this equation is associated with a WKI-type spectral problem, like the equation studied in \cite{KIW} does, which describes the propagation of a loop along a stretched rope. The electric field in an optical fiber, however, is not a stretched rope, and the SPE solution for an ultra-short light pulse we are looking for must be neither multi-valued nor singular one.

\begin{figure}
	\includegraphics[width=6cm]{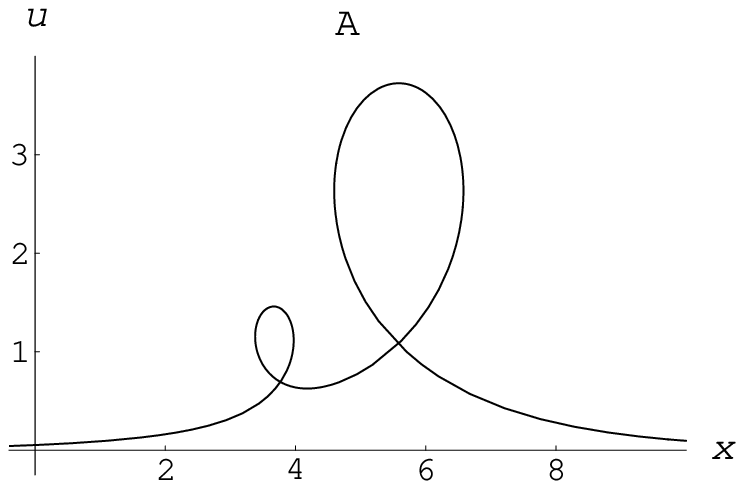}
	\includegraphics[width=6cm]{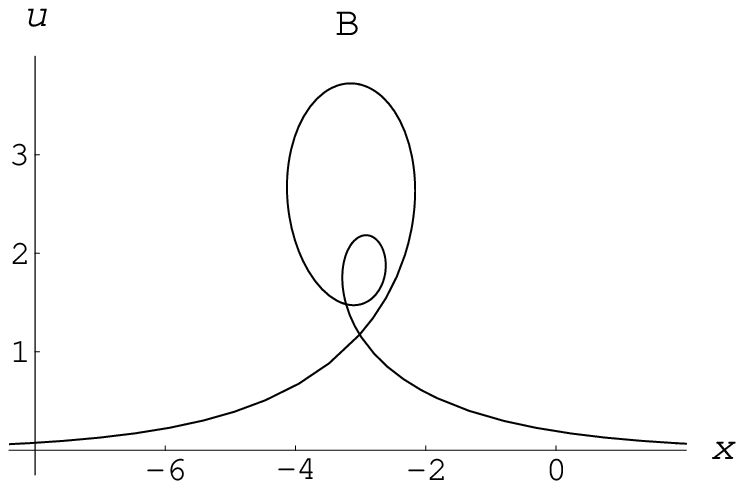}
	\caption{The two-loop solution $u(x,t)$ \eqref{2-l} of the SPE \eqref{spe} with $m = 1.2$: (A) $t = -1.6$; (B) $t = 0.9$.}
	\label{fig2}
\end{figure}

\begin{figure}
	\includegraphics[width=6cm]{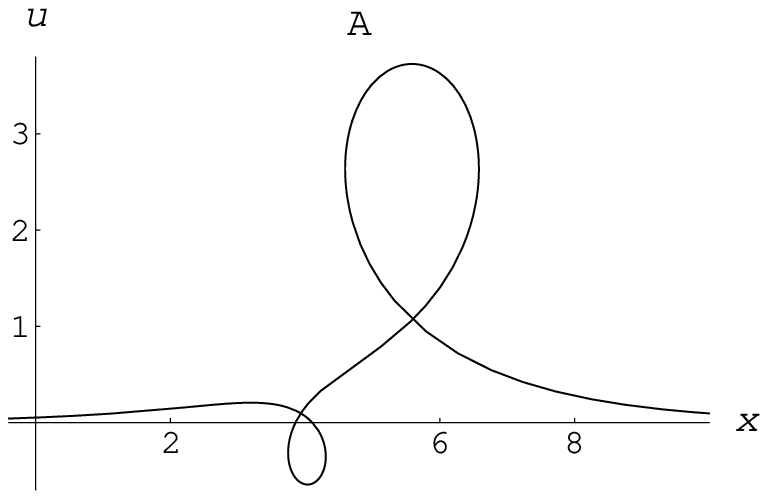}
	\includegraphics[width=6cm]{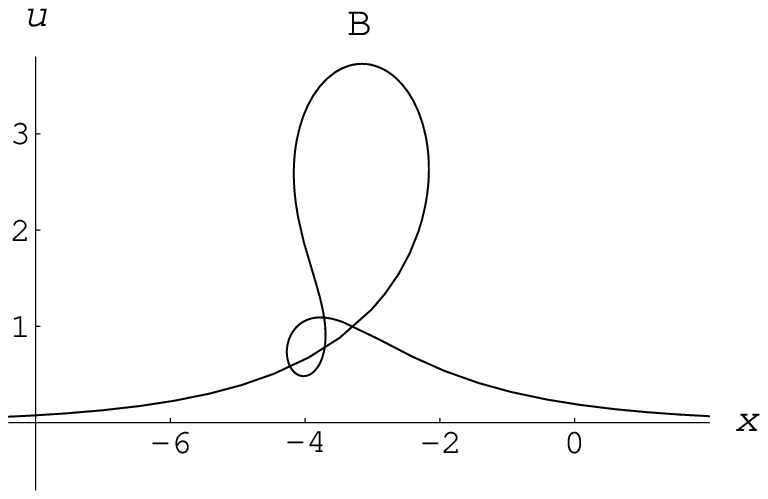}
	\caption{The loop-antiloop solution $u(x,t)$ \eqref{l-a} of the SPE \eqref{spe} with $m = 1.2$: (A) $t = -1.6$; (B) $t = 0.9$.}
	\label{fig3}
\end{figure}

It is quite obvious that any solution of the SGE, which contains an asymptotically free kink (or antikink) at large $t$, corresponds via \eqref{tru}--\eqref{trx} to a solution of the SPE containing a loop (resp., antiloop). Let us illustrate this with the two-kink solution
\begin{equation}
	z = - 4 \arctan \left( \dfrac{m \cosh \psi}{n \sinh \phi} \right)
	\label{2-k}
\end{equation}
and the kink--antikink solution
\begin{equation}
	z = - 4 \arctan \left( \dfrac{m \sinh \psi}{n \cosh \phi} \right)
	\label{k-a}
\end{equation}
of the SGE \eqref{sge} \cite{L}, where
\begin{equation}
	\phi = m ( y + t ) , \quad \psi = n ( y - t ) , \quad n = \sqrt{m^2 - 1} ,
	\label{not1}
\end{equation}
$m$ is a real parameter, $m > 1$. Applying the transformation \eqref{tru}--\eqref{trx} to the expressions \eqref{2-k} and \eqref{k-a}, and using the Mathematica system \cite{W} to do the required symbolic integration and simplification, we get, respectively, the two-loop solution
\begin{equation}
	\begin{split}
	u & = 4 m n \dfrac{m \cosh \psi \cosh \phi + n \sinh \psi \sinh \phi}{m^2 \cosh^2 \psi + n^2 \sinh^2 \phi} , \\
	x & = y + 2 m n \dfrac{m \sinh 2 \psi - n \sinh 2 \phi}{m^2 \cosh^2 \psi + n^2 \sinh^2 \phi}
	\end{split}
	\label{2-l}
\end{equation}
and the loop--antiloop solution
\begin{equation}
	\begin{split}
	u & = 4 m n \dfrac{m \sinh \psi \sinh \phi + n \cosh \psi \cosh \phi}{m^2 \sinh^2 \psi + n^2 \cosh^2 \phi} , \\
	x & = y + 2 m n \dfrac{m \sinh 2 \psi - n \sinh 2 \phi}{m^2 \sinh^2 \psi + n^2 \cosh^2 \phi}
	\end{split}
	\label{l-a}
\end{equation}
of the SPE \eqref{spe}, with the notations \eqref{not1}. Since these solutions \eqref{2-l} and \eqref{l-a}, shown in Figures \ref{fig2} and \ref{fig3}, are manifestly multi-valued, they cannot represent light pulses in optical fibers.

It is easy to find out why and when the transformation \eqref{tru}--\eqref{trx}, being applied to a solution of the SGE, yields a solution of the SPE with a singularity. From the relations \eqref{tru} and \eqref{trx} we have
\begin{equation}
	u_x ( x , t ) = \tan z ( y , t ) .
	\label{rel}
\end{equation}
Then it immediately follows from \eqref{rel} that a target solution $u(x,t)$ of the SPE can be free from singularities only if the corresponding source solution $z(y,t)$ of the SGE nowhere reaches any of the values $\left(k + \tfrac{1}{2} \right) \pi$ with $k = 0, \pm 1, \pm 2, \dotsc$. Of course, the solutions \eqref{kink}, \eqref{2-k} and \eqref{k-a} which we have already used, as well as all other solutions of the SGE which contain asymptotically free kinks or antikinks at large $t$, do not satisfy this requirement. For this reason, if we want to obtain nonsingular solutions of the SPE, we have to use those solutions of the SGE which correspond to bound states of kinks and antikinks.

Let us take the breather solution of the SGE \eqref{sge}, which is known to be the bound kink--antikink state \cite{L}:
\begin{equation}
	z = - 4 \arctan \left( \dfrac{m \sin \psi}{n \cosh \phi} \right) ,
	\label{br}
\end{equation}
where
\begin{equation}
	\phi = m ( y + t ) , \quad \psi = n ( y - t ) , \quad n = \sqrt{1 - m^2} ,
	\label{not2}
\end{equation}
$m$ is a real parameter, $0 < m < 1$. Applying the transformation \eqref{tru}--\eqref{trx} to the expression \eqref{br}, we get the following solution of the SPE \eqref{spe}:
\begin{equation}
	\begin{split}
	u & = 4 m n \dfrac{m \sin \psi \sinh \phi + n \cos \psi \cosh \phi}{m^2 \sin^2 \psi + n^2 \cosh^2 \phi} , \\
	x & = y + 2 m n \dfrac{m \sin 2 \psi - n \sinh 2 \phi}{m^2 \sin^2 \psi + n^2 \cosh^2 \phi} ,
	\end{split}
	\label{ps}
\end{equation}
with the same notations \eqref{not2}. Note that we used the breather solution of the SGE in the form \eqref{br} simplified by a Lorentz transformation and shifts of $y$ and $t$, and the constant of integration in $x$ \eqref{ps} has been fixed. If necessary, one can generalize the solution $u(x,t)$ \eqref{ps} by the scale transformation \eqref{str} and shifts of $x$ and $t$.

\begin{figure}
	\includegraphics[width=6cm]{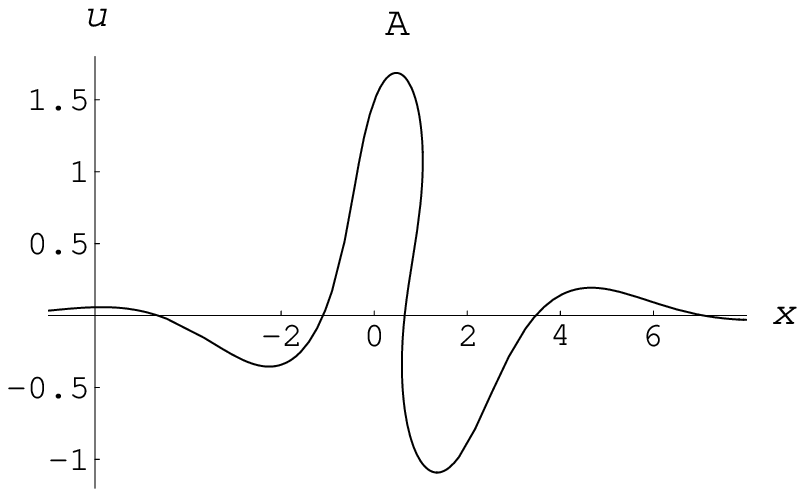}
	\includegraphics[width=6cm]{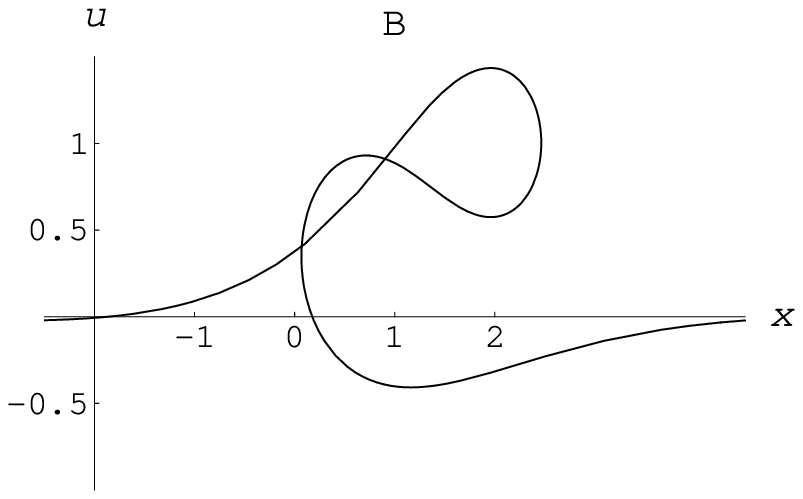}
	\caption{The pulse solution $u(x,t)$ \eqref{ps} of the SPE \eqref{spe} with $m > m_{\mathrm{cr}}$: (A) $m = 0.5$, $t = -0.6$; (B) $m = 0.8$, $t = -0.9$.}
	\label{fig4}
\end{figure}

The solution \eqref{ps} of the SPE \eqref{spe} is the central result of this paper, and hereinafter it will be referred to as the pulse solution. This name does not mean, however, that the expressions \eqref{ps} represent a nonsingular pulse for any value of the parameter $m$, $0 < m < 1$. The function $z(y,t)$ \eqref{br} satisfies the condition $- \tfrac{\pi}{2} < z < \tfrac{\pi}{2} $ for all values of $y$ and $t$ only if $m < m_{\mathrm{cr}}$, where
\begin{equation}
	m_{\mathrm{cr}} = \sin \tfrac{\pi}{8} \approx 0.383 .
	\label{mcr}
\end{equation}
Therefore, when the value of $m$ reaches or exceeds the critical value $m_{\mathrm{cr}}$ \eqref{mcr}, the singularities $u_x \to \pm \infty$ appear in the function $u(x,t)$ \eqref{ps} due to the relation \eqref{rel}. Furthermore, in the overcritical case, when $m > m_{\mathrm{cr}}$, the pulse solution \eqref{ps} is multi-valued, as shown in Figure~\ref{fig4}.

\begin{figure}
	\includegraphics[width=6cm]{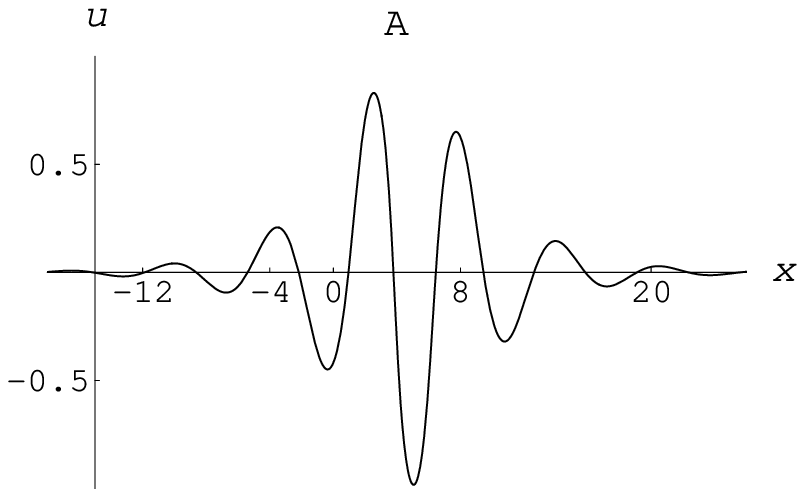}
	\includegraphics[width=6cm]{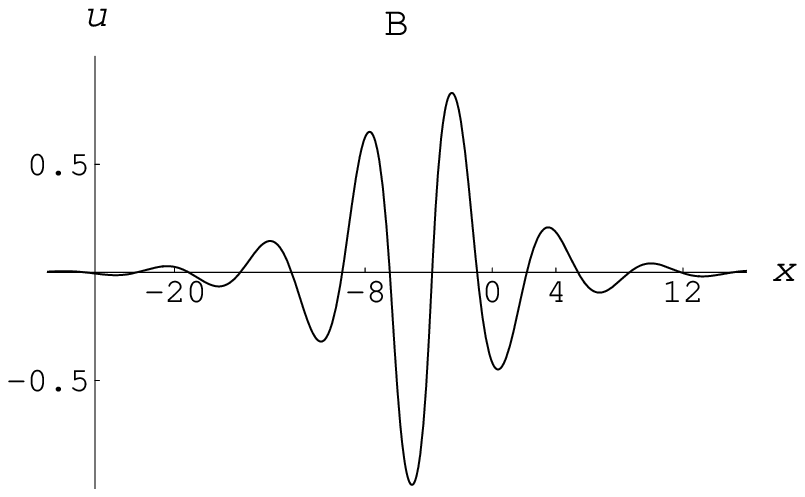}
	\caption{The pulse solution $u(x,t)$ \eqref{ps} of the SPE \eqref{spe} with $m = 0.25 < m_{\mathrm{cr}}$: (A) $t = -4.5$; (B) $t = 4.5$.}
	\label{fig5}
\end{figure}

\begin{figure}
	\includegraphics[width=6cm]{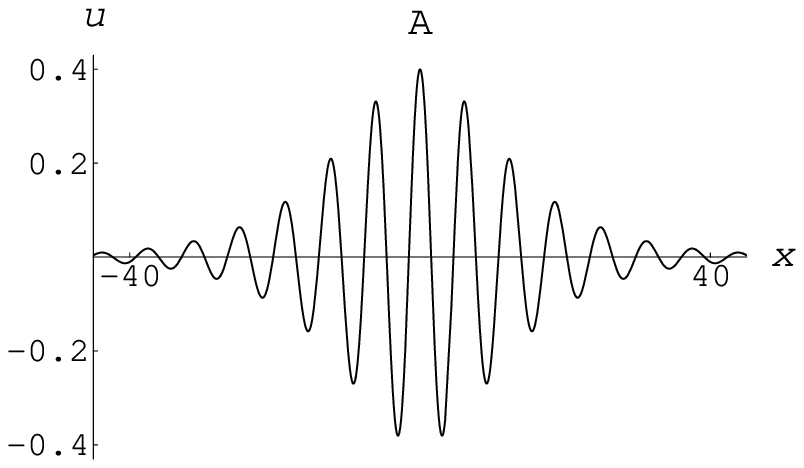}
	\includegraphics[width=6cm]{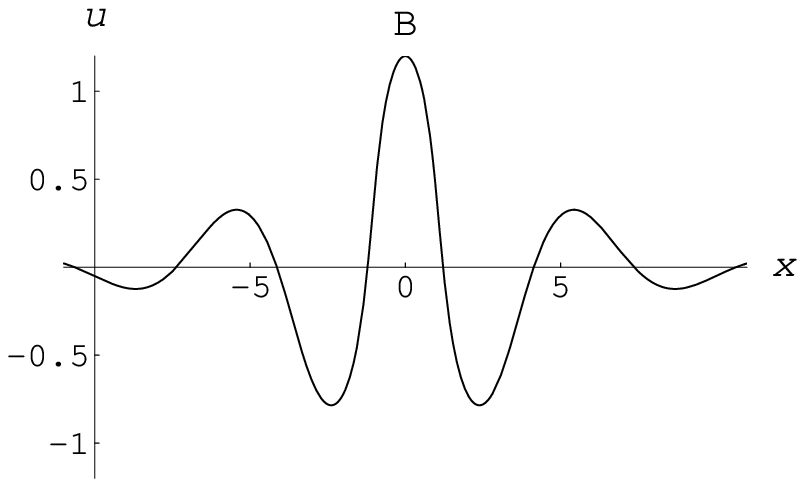}
	\caption{The pulse solution $u(x,t)$ \eqref{ps} of the SPE \eqref{spe} with $m < m_{\mathrm{cr}}$, at $t = 0$: (A) $m = 0.1$; (B) $m = 0.3$.}
	\label{fig6}
\end{figure}

In the undercritical case, when $m < m_{\mathrm{cr}}$, the pulse solution \eqref{ps} represents a single-valued nonsingular pulse or, more precisely, a wave packet. Its typical shape is shown in Figure~\ref{fig5}.
The envelope curve of this wave packet is determined by the hyperbolic functions in the expressions \eqref{ps} and moves from the right to the left with unit speed, whereas the oscillatory component is determined by the trigonometric functions in \eqref{ps} and moves from the left to the right, also with unit speed. The smaller is the value of the parameter $m$ in the pulse solution \eqref{ps}, the larger is the number of oscillations in the represented wave packet, as shown in Figure~\ref{fig6}.
When the value of $m$ is small, the pulse solution \eqref{ps} can be approximated as
\begin{equation}
	u \approx 4 m \cos ( y - t ) \operatorname{sech} \bigl( m ( y + t ) \bigr) , \quad x \approx y ,
	\label{sm}
\end{equation}
and the shape of the SPE pulse is similar to that of the NLSE soliton: the sech-shaped envelope modulates the cos-shaped wave (see Figure~\ref{fig6}A). On the other hand, very short pulses of the SPE, like the one shown in Figure~\ref{fig6}B, correspond to near-critical values of $m$, for which the approximation \eqref{sm} does not work fine and one has to use the exact expressions \eqref{ps}. The ultra-short pulses represented by the pulse solution \eqref{ps} with near-critical values of $m$ can be as short as approximately three cycles of their central frequency.

\begin{figure}
	\includegraphics[width=6cm]{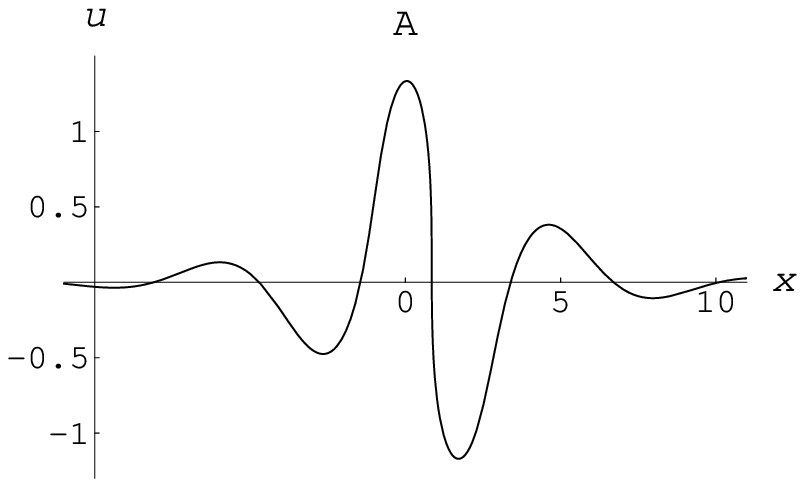}
	\includegraphics[width=6cm]{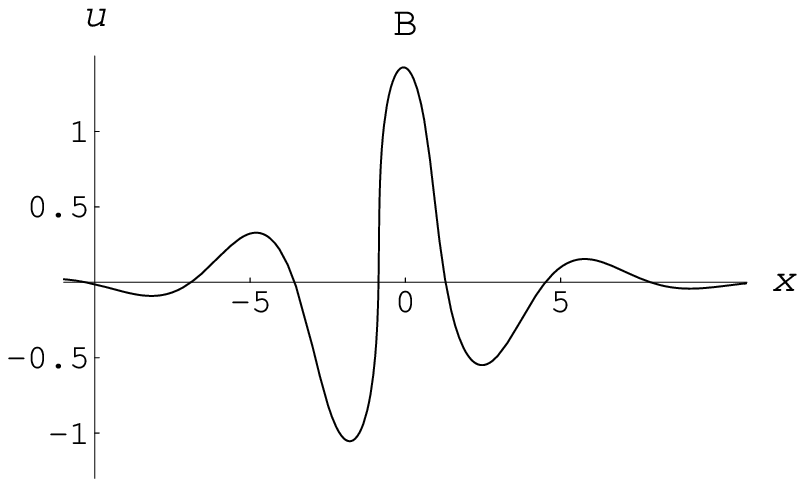}
	\caption{The pulse solution $u(x,t)$ \eqref{ps} of the SPE \eqref{spe} with $m = m_{\mathrm{cr}} \approx 0.383$: (A) $t = -0.7$; (B) $t = 0.5$.}
	\label{fig7}
\end{figure}

In the critical case, when $m = m_{\mathrm{cr}} = \sin \tfrac{\pi}{8}$, the pulse solution \eqref{ps} still remains single-valued, but the singularities $u_x \to \pm \infty$ appear in it. This critical wave packet, or the shortest single-valued pulse of the SPE which we were able to find using the transformation \eqref{tru}--\eqref{trx} between the SPE \eqref{spe} and the SGE \eqref{sge}, is shown in Figure~\ref{fig7}.
At present, we do not know whether the SPE can describe a shorter wave packet. It may well be that not, and that a modification of the model equation is necessary for extremely short optical pulses in nonlinear media, like the one proposed very recently in \cite{CS}.

In conclusion, we can formulate the main result of the present paper as follows. We have found an exact nonsingular solitary wave solution of the Sch\"{a}fer--Wayne short pulse equation. This pulse solution was derived from the breather solution of the sine-Gordon equation by means of a transformation between these two integrable equations. The ultra-short wave packets represented by the obtained pulse solution can be as short as approximately three cycles of their central frequency.

\end{document}